\documentclass{appolb}
\usepackage{graphicx}
\usepackage[symbol]{footmisc}

\begin{document}

\title{Dedicated $\Delta$E-E detector system for searching long lived heaviest nuclei deposited in scintillators%
\thanks{Presented at Zakopane Conference on Nulear Physics 2018}%
}
\author{
	K. Zelga$^a$\footnote{kamila.zelga@gmail.com}, 
	Z. Majka$^a$, 
	R. P\l{}aneta$^a$, 
	Z. Sosin$^a$\footnote{Deceased}, 
	A. Wieloch$^a$, 
	M. Adamczyk$^a$, 
	K. \L{}ojek$^a$,
	M. Barbui$^b$, 
	S. Wuenschel$^b$, 
	K. Hagel$^b$, 
	X. Cao$^b$, 
	J. Natowitz$^b$, 
	R. Wada$^b$, 
	G. Giuliani$^b$,
	E-J. Kim$^c$,
	H. Zheng$^d$,
	S. Kowalski$^e$,
\address{
	$^a$ M. Smoluchowski Institute of Physics, Jagiellonian University, \L{}ojasiewicza 11, 30-348 Krak\'{o}w, Poland
	\\$^b$ Cyclotron Institute, Texas A\&M University, College Station, Texas 77843, USA
	\\$^c$ Division of Science Education, Chonbuk National University, Jeonju, 54896, Korea
	\\$^d$ School of Physics and Information Technology, Shaanxi Normal University, Xi’an 710119, China
	\\$^e$ Institute of Physics, University of Silesia, 75 Pu\l{}ku Piechoty 1, 41-500 Chorz\'{o}w, Poland}
}
\maketitle
\begin{abstract}
We present a dedicated experimental setup which is currently used to search for long lived super heavy elements (SHE) implanted in catcher scintillators which were irradiated by reaction products of $^{197}$Au (7.5 A.MeV) projectile and $^{232}$Th target collisions during our experiment performed at Cyclotrone Institute, Texas A$\&$M University in 2015. The built-in novel measuring apparatus consists of $\Delta$E-E detector pairs which are able to register $\alpha$ or spontaneous fission (SF) decays of heavy reaction products deposited in the scintillators. Their unique feature is that the examined scintillators are at the same time $\Delta$E part of each of $\Delta$E-E detector while E part is a silicon detector. Our apparatus is dedicated to search for SHEs which have a lifetime of a year till tens of years. Results of commissioning tests of our setup are presented.
\end{abstract}

\section{Introduction}
The search for the island of stability of SHEs is one of the most challenging problems in nuclear physics. All already discovered SHE isotopes, mainly in complete fusion reactions, are characterized by lifetimes of only a few minutes or shorter \cite{Ogan:06:1,Hofm:02:1,Mori:04:1}. However many attempts to search for stable (lifetime of billions of years) SHEs in Nature have not yet yielded positive results \cite{Kors:15:1}. Along the lifetime axis of heaviest elements, there is a time region of the order of years that is accessible for experimentalists, but so far only briefly explored \cite{Gagg:80:1}.

In recent years, scientists from Jagiellonian University and Texas A\&M University (TAMU), tested the multi-nuclen transfer reaction as a way to create SHEs \cite{Majk:18:1,Wuen:18:1}. One of the studied reactions was $^{179}$Au (7.5 A.MeV) projectile on $^{232}$Th target of 12 mg/cm$^2$ thickness and 3${\cdot}$10$^{15}$ of Au beam ions were delivered to the target. Measurements were done in the Cyclotron Institute of TAMU in 2015. The detector, based on the BC-418 plastic scintillators, prepared for the experiment, was built in such a way that segmented 63 plastic scintillators played the role of an active catcher (AC), into which the reaction products and SHEs were implanted. The aim of this experiment was to search for short-lived SHEs with life times from nanoseconds to microseconds.

In our current project, we constructed a dedicated detection apparatus, which is a simple multi-element system to search for long lived SHEs (LLSHEs) with life times of years to tens of years, deposited in the scintillators of the AC.

The article is organized as follows: in sec. \ref{setup} we describe the idea and the construction of the apparatus then in sec. \ref{results} we present some results of measurements and finally a short discussion is given in sec. \ref{summary}.

\section{Experimental setup}
\label{setup}
\begin{figure}[htb]
\centerline{%
\includegraphics[width=12.5cm]{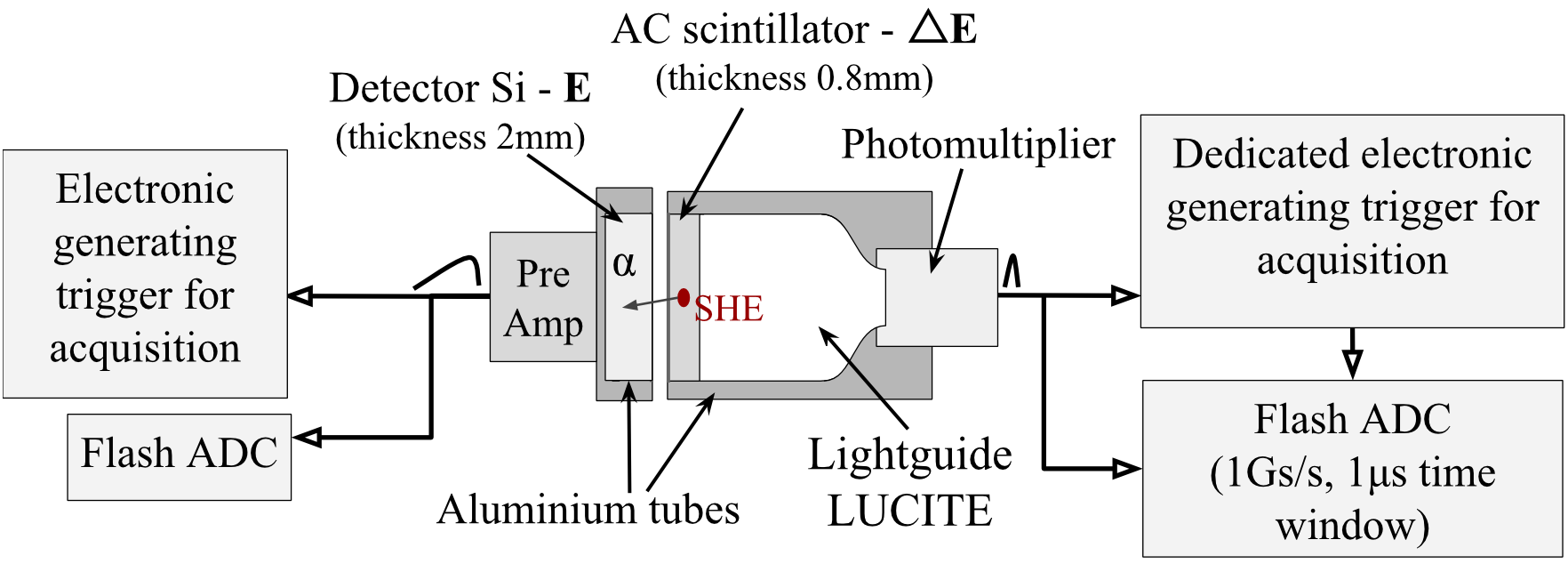}}
\caption{Scheme of $\Delta$E-E detector. See text for details.}
\label{Fig:1}
\end{figure}

The idea of the apparatus is based on the registration of $\alpha$/SF decays of LLSHEs, possibly implanted in the AC scintillators. To search for such decays, $\Delta$E-E detectors were constructed. A scheme of one such pair is presented in Fig. \ref{Fig:1} where $\Delta$E is the AC scintillator and E is lithium drifted silicon detector (Si). The AC detection element ($\Delta$E) is attached to a light guide placed inside a cavity to accumulate light on a photomultiplier tube (R9880U-110). Front of the scintillator is covered by thin aluminium foil, below 1 $\mu$m thickness. The E part is connected to the charge pre-amplifier which is placed very close to the detector. Both detectors are placed in the air facing each other with a small gap between them (less then 1 mm).

Based on the $\Delta$E-E pairs a simple multi-element system was built. The whole setup consists two walls of detectors. A wall of 8 Si detectors provide information about energy of the registered $\alpha$/SF particle, while the second wall includes 8 AC scintillators, which are explored scintillators, and at the same time $\Delta$E detectors. Both walls are placed on movable rails, to adjust the $\Delta$E-E mutual positions.

The background is a serious problem in the search for very rare events which are decays of SHEs. It is composed of natural radiation from the environment from various radioactive decay chains (thorium, uranium-radium, etc.) and cosmic radiation. Natural radiation contains $\alpha$, $\beta$, $\gamma$ particles of which only $\alpha$ particles can mimic the part of the genetic chain from the decay of SHE. Fortunately, the energy of $\alpha$ emitted at the beginning of a genetic chain in the decay of SHE, reaches the value of 10 MeV and higher \cite{Bend:13:1}, while the highest energy of $\alpha$ particles coming from natural radiation is 8.99 MeV (decay of $^{217}$Ra).
Estimated depths of implanted SHEs in AC scintillators are a few microns \cite{Wuen:18:1}. Some of the $\alpha$ particles from the decay of SHEs will be emitted in the direction of the E detector, leaving only a small part of energy in the AC scintillator ($\Delta$E) and in the small gap of the air between $\Delta$E-E, while the greater part of their energy will be registered by the Si detector. At the same time, this scintillator is an anti-coincidence shield against cosmic radiation, because its actual thickness is 0.8 mm, while energy deposited by such radiation in Si detector is not higher than a few MeV. In the case of SF from SHE decays, the energy registered in both detectors should be high. Estimated effective geometrical efficiency of the $\Delta$E-E pair for interesting events is not higher then 20$\%$ of a 4$\pi$ solid angle.
 
Signals from any of the $\Delta$E or E detector pairs generated by the registered  particles (e.g., $\alpha$/SF from SHE decay in the $\Delta$E scintillator) are split and sent into analog and digital logic branches of the electronics (see Fig. \ref{Fig:1}). To produce triggers from signals of AC scintillators dedicated electronics were used (see Fig. 3a in \cite{Majk:18:1}), while in the case of Si detectors standard fan-in fan-out modules (Caen v925) with discrimination capability were applied. The signals themselves were recorded as a wave forms using the Caen FADC V1742 digitizer modules. These modules were set to a sampling rate of 1 Gs/s in the case of Si detectors and 1 and 5 Gs/s in the case of the scintillator detectors with a 1024 points buffer. Inspection of the recorded signal waveforms saved on HDD drive allow us to distinguish in further analysis the real physical signal from an artificial disturbance. All detectors were calibrated in the vacuum with sources of $^{241}$Am (which emits $\alpha$) and $^{252}$Cf (which emits $\alpha$ as well as SF fragments). Energy resolution deduced from the calibration procedure was 1\% in case of Si detectors and around 30\% in the case of AC scintillators. 

In the next section we present measurement results to discuss the performance of our detection system. 

\section{Presentation of results}
\label{results}

\begin{figure}[htb]
\centerline{%
\includegraphics[width=11cm]{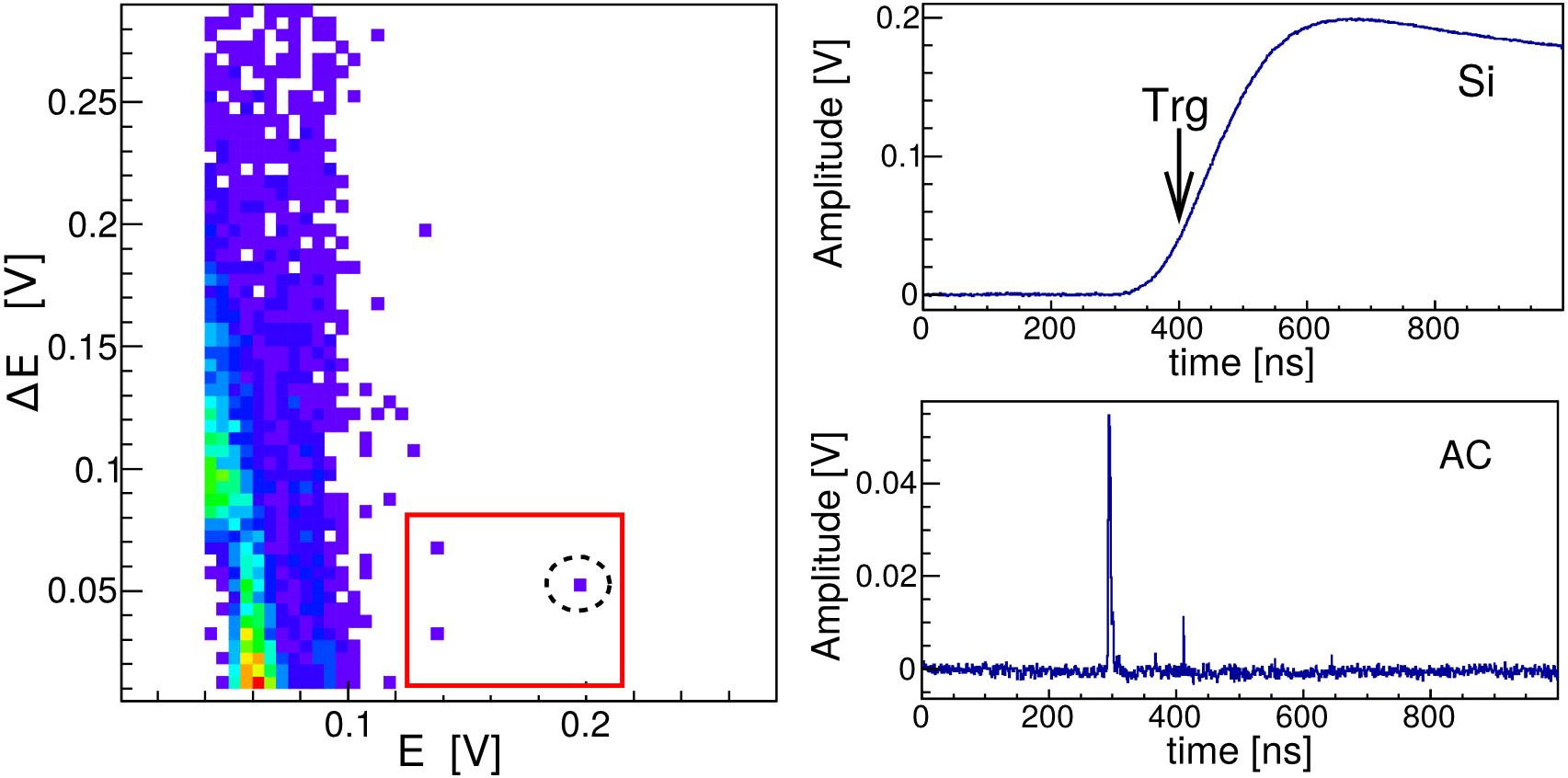}}
\caption{Left side: $\Delta$E-E map for Si-AC pair. Right side: pulses from Si and AC detectors for the event marked inside dashed circle on the $\Delta$E-E map.}
\label{Fig:3}
\end{figure}

The most convenient way to search for interesting events (candidates for SHE decays) is an anlysis of two-dimensional maps of $\Delta$E-E energies. In left panel of Fig. \ref{Fig:3} we present a typical map for the AC scintillator - Si detector pair, received after 25 days of continuous measurement. A rectangular area on the map is showing expected localization of interesting events. 
The heights of this rectangle is estimated from the observation that the SHEs should be implanted in the AC scintillator at depths of several microns \cite{Wuen:18:1}, hence it is assumed to be around 2 MeV (80 mV). The base of the rectangle is chosen to be in the range of $\alpha$ energies expected for decays of heaviest SHEs, i.e. 10-18 MeV \cite{Bend:13:1}. 
In this area one can see three interesting events and each one of those events was directly inspected to check its physical origin. On the right panel of this figure we present an example signal of the event marked by dashed circle. We see a relatively slow well-defined silicon pulse which confirms the detection of a charged particle. Its energy is 16 MeV. In the AC scintillator the pulse is very fast, with a duration of 10 ns and energy around 1 MeV. The spike of the 10 mV amplitude seen on the right of the pulse is commonly present in the acquisition and its origin is the digital nature of the acquisition electronic system. As one can see, the fast pulse from the AC scintillator is in place when a slower pulse from the Si detector starts. Both pulses are properly located with respect to the arrival of the ACQ trigger (a vertical arrow on the figure) which confirms that presented pulses from $\Delta$E and E detectors were produced by the same particle. 
 
\begin{figure}[htb]
\centerline{%
\includegraphics[width=12.5cm]{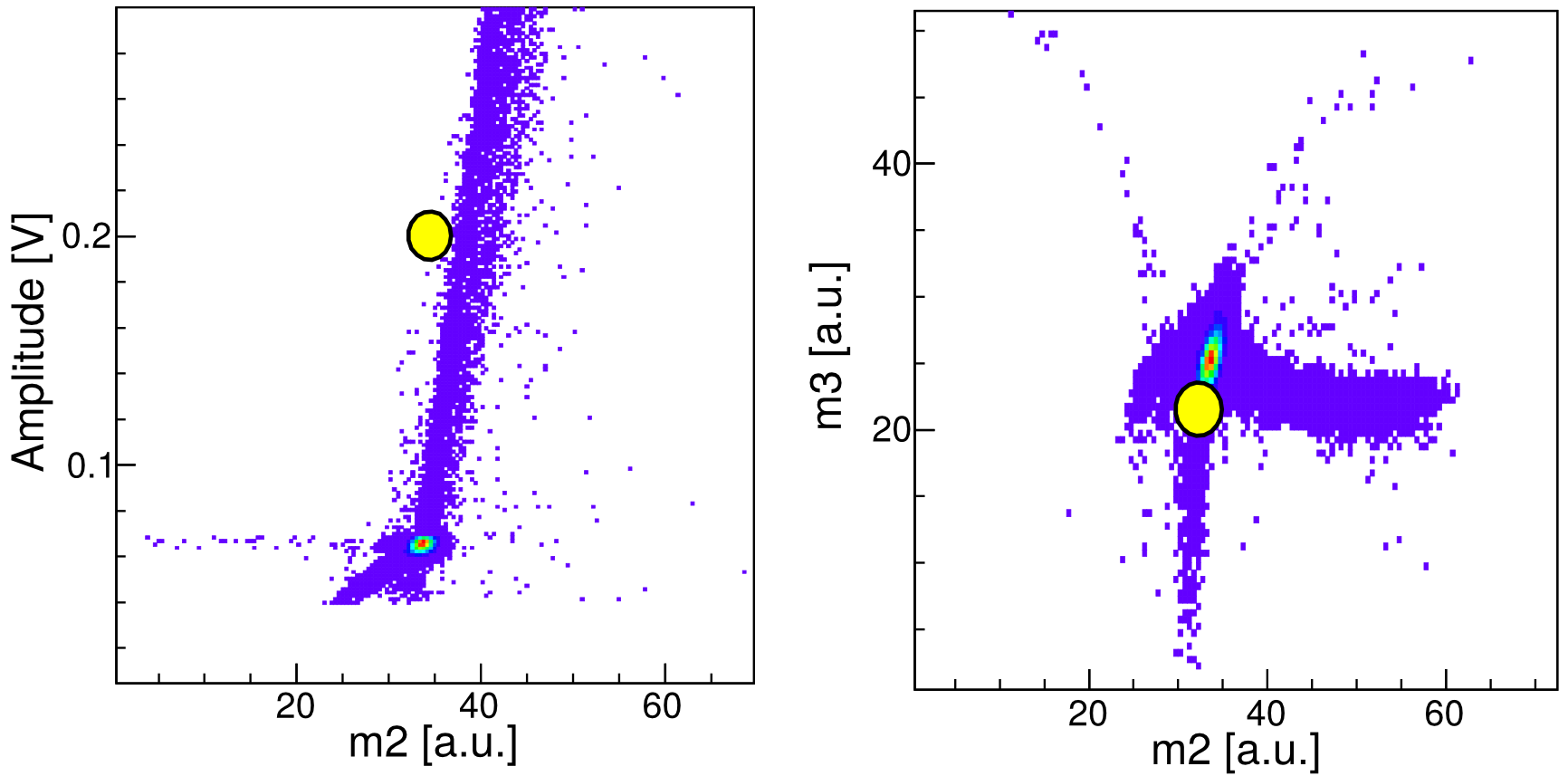}}
\caption{Left side: map of charge pulse amplitude vs m$_2$ of current pulse. Right side: map of m$_3$ vs m$_2$ of current pulse. Measurement for Cf source.}
\label{Fig:4}
\end{figure}

The recorded pulses can be further analyzed by using digital signal processing (DSP) tools. This is especially straightforward for signals from the Si detector where DSP methods frequently enable particle identification. For this, it is useful to get characteristics of the Si pulse such as its rise time, its amplitude or the amplitude of the current pulse, which is a derivative of the original one and also the second (m$_2$) and third (m$_3$) moments of the current pulse \cite{Barl:09:1}. In Fig. \ref{Fig:4} we present two examples of a DSP applied to the data collected from calibration measurements with the $^{252}$Cf source. On the right side a 2-dimensional map for the charge (original) pulse amplitudes versus m$_2$ of the current pulses is shown. The left figure presents the third moment of current pulse versus m$_2$. The maximum (yellow and light blue area in online version) seen on both maps corresponds to 6.1 MeV $\alpha$ particles emitted by the californium, while the dark tail (blue colour in online version) spanning upwards on the left panel and to the right on the second panel represent the SF events of our source. This type of map can be treated as a way of showing patterns validating interesting events, e.g. event presented on the previous figure, which is indicated here as a solid circle (yellow colour in online version).

In the case of AC scintillators DSP tools are not so obvious. This is due to the very fast pulse registered from the scintillators and this needs a new DSP approach to extract more detailed information about particles that produced a pulse.

\section{Summary and conclusions}
\label{summary}

In our paper we have presented a dedicated detection apparatus constructed to search for LLSHEs created in heavy ion collisions and deposited in scintillator material. Presented results show that our detectors are capable of recording interesting events using coincidence techniques on an $\Delta$E-E map. Presently we are collecting more data that will be carefully analysed using digital signal processing to distinguish more precisely $\alpha$ particles from the SF fragments. 

We appreciate Dr. Gina Agarwal, McMaster University, Canada, for helpful editing suggestions.

This work is supported by DSC 2018 grant at WFAIS UJ - financed by the Ministry of Science and Higher Education 7150/E-338/M/2018, No. K/DSC/005314, by the National Science Center in Poland, contract no. UMO-2012/04/A/ST2/00082, by the U.S. Department of Energy under Grant No. DE-FG03-93ER40773 and by the Robert A. Welch Foundation under Grant A0330.

\end{document}